\documentclass{aa} 

\usepackage{graphicx}
\usepackage{txfonts}
\usepackage[citecolor=blue, linkcolor=blue, urlcolor = black, colorlinks = true]{hyperref}

\usepackage{graphicx}	
\usepackage{amsmath}	
\usepackage{amssymb}	
\usepackage{xspace}

\usepackage{siunitx}
\usepackage{comment}
\usepackage{textcomp}
\usepackage{bm}
\usepackage{lscape}
\usepackage{xcolor}
\usepackage{natbib}
\usepackage{subcaption}

\newcommand{\Msun}{\,M$_{\odot}$\xspace}

\newcommand{\percent}{~per~cent\xspace}

\newcommand{\conny}[1]{#1}
\newcommand{\edit}[1]{#1}

\newcommand{\gyre}{\texttt{GYRE}\xspace}
\newcommand{\topc}{\texttt{TOP}\xspace}
\newcommand{\ester}{\texttt{ESTER}\xspace}
\newcommand{\storm}{\texttt{StORM}\xspace}
\newcommand{\cpd}{$\,{\rm d}^{-1}$\xspace}

\usepackage{hyperref}
\begin{document}

   \title{Is a 1D perturbative method sufficient for asteroseismic modelling of $\beta$~Cephei pulsators? }
   \subtitle{Implications for measurements of rotation and internal magnetic fields}
   \author{J.~S.~G. Mombarg\inst{1}
          \and
          V. Vanlaer\inst{2}
          \and 
          S.~B. Das\inst{3,4}
          \and 
          M. Rieutord\inst{5}
          \and
          C. Aerts\inst{2,6,7} 
          \and 
          L. Bugnet\inst{3}
          \and 
          S. Mathis\inst{1}
          \and \\
          D. R. Reese\inst{8}
          \and
          J. Ballot\inst{5}
          }

   \institute{Universit\'e Paris-Saclay, Universit\'e de Paris, Sorbonne Paris Cit\'e, CEA, CNRS, AIM, 91191 Gif-sur-Yvette, France\\
              \email{joey.mombarg[@]cea.fr}
              \and 
              {Institute of Astronomy, KU Leuven, Celestijnenlaan 200D, B-3001 Leuven, Belgium} 
              \and 
              {Institute of Science and Technology Austria (ISTA), Am Campus 1, Klosterneuburg, Austria}
               \and 
              {Center for Astrophysics | Harvard \& Smithsonian, 60 Garden Street, Cambridge, MA 02138, USA}
              \and 
              {IRAP, Université de Toulouse, CNRS, UPS, CNES, 14 avenue Édouard Belin, F-31400 Toulouse, France}
              \and
              {Department of Astrophysics, IMAPP, Radboud University Nijmegen, PO Box 9010, 6500 GL Nijmegen, The Netherlands}
              \and
              {Max Planck Institute for Astronomy, Königstuhl 17, 69117, Heidelberg, Germany}
              \and
              {LIRA, Observatoire de Paris, Université PSL, Sorbonne Université, Université Paris Cité, CY Cergy Paris Université, CNRS, 92190 Meudon, France}             
        }

   \date{Received September 15 2025; accepted November 10 2025}
\titlerunning{Comparing 1D and 2D asteroseismology in the perturbative regime}
\authorrunning{Mombarg et al.}
 
  \abstract
   {Asymmetries in the observed rotational splittings of a multiplet contain information about the star's rotation profile and internal magnetic field. Moreover, the frequency regularities of multiplets can be used for mode identification. However, to exploit this information, highly accurate theoretical predictions are needed.
   }
   {We aim to quantify the difference in the predicted mode asymmetries between a 1D perturbative method, and a 2D method that includes a 2D stellar structure model, which takes rotation into account. We then put these differences between 1D and 2D methods in the context of asteroseismic measurements of internal magnetic fields. We only focus on the asymmetries and not on possible additional frequency peaks that can arise when the magnetic and rotation axis are misaligned.  }
   {We couple the 1D pulsation codes \gyre and \storm to the 2D stellar structure code \ester and compare the oscillation predictions with the results from the 2D \topc pulsation code. We focus on zero-age main-sequence models representative of rotating $\beta$~Cephei pulsators, going up to 20 per cent of the critical Keplerian rotation rate. Specifically, we investigate low-radial order gravity and pressure modes. }
   {We find a generally good agreement between the oscillation frequencies resulting from the 1D and 2D pulsation codes. We report differences in predicted mode multiplet asymmetries mostly below $0.06\,{\rm d^{-1}}$. Since the magnetic asymmetries are small compared to the differences in the rotational asymmetries resulting from the 1D and 2D predictions, accurate measurements of the magnetic field are in most cases challenging.  }
   {Differences in the predicted mode asymmetries of a rotating star between 1D perturbative methods and 2D non-perturbative methods can greatly hinder accurate measurements of internal magnetic fields in main-sequence pulsators with low-order modes. Nevertheless, reasonably accurate measurements could be possible with $n_{\rm pg} \ge 2$ modes if the internal rotation is roughly below 10\percent of the Keplerian critical rotation frequency for (aligned) magnetic fields on the order of a few hundred kG. While the differences between the 1D and 2D frequency predictions are mostly too large for internal magnetic field detections, the rotational asymmetries predicted by \storm are in general accurate enough for asteroseismic modelling of the stellar rotation in main-sequence stars with identified low-order modes.   }


   \maketitle
%
\section{Introduction}
In the absence of stellar rotation or magnetism, the frequencies of stellar pulsation modes of equal spherical degree ($\ell$) and radial order ($n_{\rm pg}$) are degenerate for different azimuthal orders ($m$). This degeneracy in frequency is lifted by rotation, splitting a single frequency into a multiplet of $2\ell + 1$ components. 
Up to first order in the rotation frequency $\Omega$,
the splitting
is symmetric, namely $\omega_m = \omega_0 + m(1 - C_{\ell,n})\Omega$, where $\omega_m$ is the frequency in the inertial frame and $C_{\ell, n}$ is the Ledoux constant \citep{Ledoux1951}. Higher order effects also occur due to the Coriolis force and the centrifugal deformation of the mode cavities \citep[e.g.][second and third order]{saio1981, gough1990, goode1992, soufi1998,karami2008, Guo2024}. The non-zonal ($m \not= 0$) modes are more sensitive to the (rotational) equator and therefore experience a larger frequency shift compared to the zonal mode. This introduces asymmetries in the frequency splittings. Likewise, the presence of a magnetic field introduces asymmetries in the splittings \citep{Loi2020, GomesLopes2020,  bugnet2021, Mathis2021, Bugnet2022, li2022,  MathisBugnet2023, Das2020, Das2024}. Such observed asymmetries have been exploited to estimate internal magnetic field strengths in red giants where the contribution of rotation to the asymmetry can be neglected, following the framework developed by \cite{li2022} and used in follow-up studies \citep{li2023, Hatt2024}. 

For main sequence pulsators that show frequency asymmetries, such as the  $\beta$~Cephei and $\delta$~Scuti pulsators, rotation cannot simply be neglected, but we note these two classes of stars are typically in different regimes in terms of the fraction of critical rotation \citep[see e.g.][]{Huang2010}, and in terms of excited radial orders.  It is therefore crucial to have a good understanding of the effects of rotation on the asymmetries. Here, we focus on $\beta$~Cephei pulsators. Mode identification of these stars starts to become feasible on a large scale 
\conny{from the analyses of combined {\it Gaia\/} and TESS photometry
\citep[e.g.][]{DeRidder2023,HeyAerts2024,Fritzewski2025}.} In the context of predicting mode asymmetries, all studies so far have relied on one-dimensional (1D) perturbative methods. The 1D perturbative theory has been compared with 2D methods using polytropic models by \cite{Reese2006} for pressure modes, and by \cite{Ballot2010} for gravity modes. Furthermore, \cite{Burke2011} used realistic stellar structure models of stars of 1 and 2\Msun, computed with the 1D ASTEC code \citep{christensen-dalsgaard2008}, to compare the 1D and 2D methods. The validity domains of the perturbative method provided by these studies are based on a typical frequency resolution of the CoRoT mission \citep{Auvergne2009}. 

Asteroseismic modelling of real stars in 2D has been done for rapidly-rotating pulsators using both 2D steady models computed with the \ester stellar structure and evolution code \citep{EspinosaLara2013, Rieutord2016} coupled to the 2D stellar pulsation code \topc \citep{Reese2009}. Such modelling efforts typically rely on the mode visibility for the mode identification, that is, the surface pressure perturbation integrated over the visible disc \citep{Rieutord2024}. However, 2D pulsation codes have never (to our knowledge) been used to reproduce pulsation spectra of slow rotators.

In this paper, we quantify the difference in mode 
frequencies and rotationally split multiplet
asymmetries of low radial-order g and p modes predicted by the new  1D pulsation code (\storm) and a 2D pulsation code (\topc) for models representative of $\beta$~Cephei pulsators. We ignore additional frequencies that may occur in multiplets if the magnetic and rotation axis are sufficiently misaligned. This can give rise to $(2\ell +1)^2$ frequencies in the inertial frame \citep{Loi2021}, and additional frequencies in multiplets have been observed in some $\beta$~Cephei pulsators \citep[e.g.][]{ShibahashiAerts2000, Vanlaer2025-modelling, Vandersnickt2025}.

\begin{figure}[htb]
    \centering
    \includegraphics[width=0.95\linewidth]{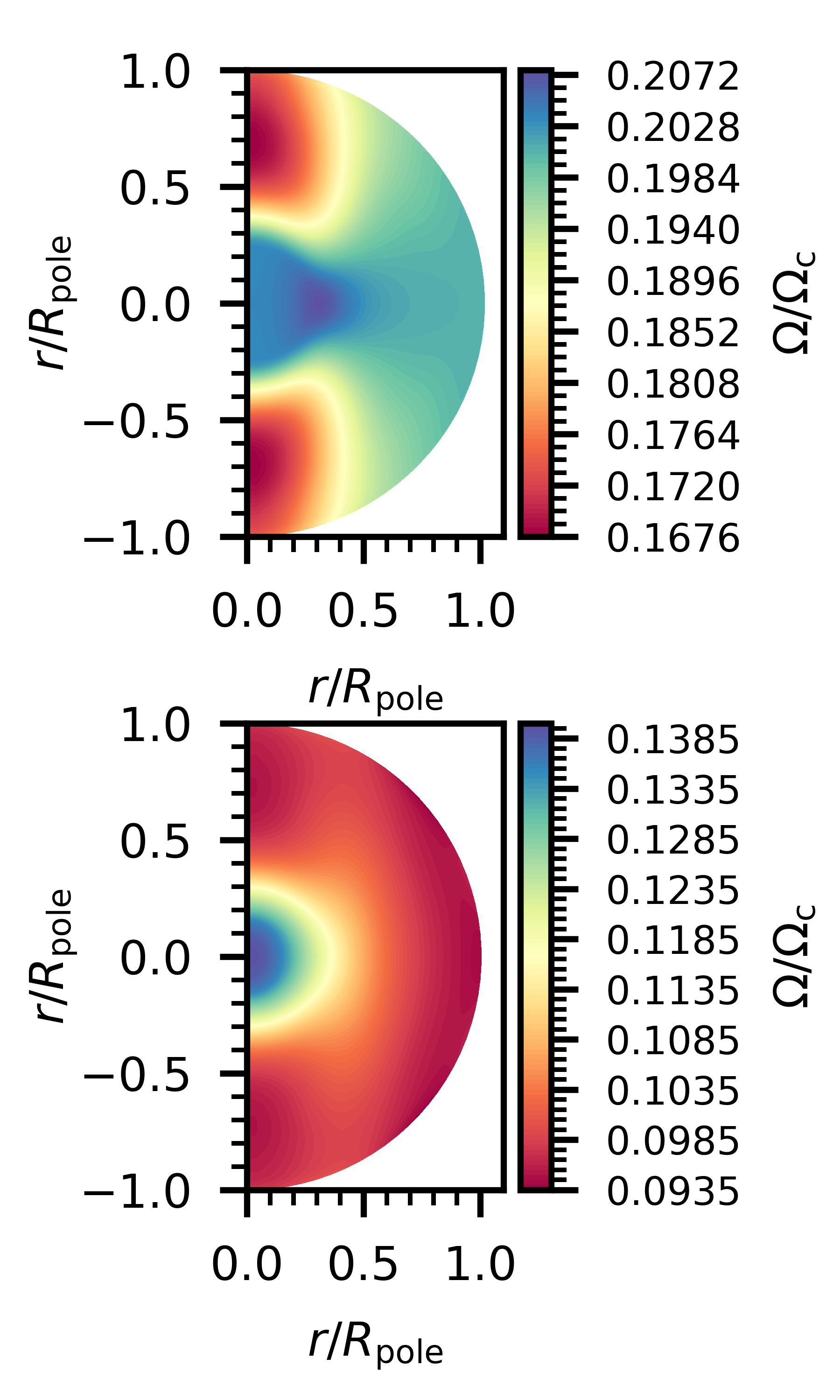}
    \caption{Differential rotation profiles of a 12\Msun stars computed with \ester at ZAMS (top panel) and $X_{\rm c} = 0.4$ (bottom panel), normalized by the critical rotation frequency at the equator. }
    \label{fig:omega_map}
\end{figure}

For the 2D case, we also use a 2D stellar structure model with a differential rotation profile resulting from the baroclinic torque, and compare this with the assumption of solid body rotation in the 1D case (see also \cite{Houdayer2023} for a comparison between a differentially rotating \ester model and a deformed uniformly rotating one). Figure~\ref{fig:omega_map} shows an example of the 2D rotation profile, both for the ZAMS model rotating at 10\percent of the critical Keplerian rotation frequency discussed in Sect.~\ref{sec:num} and for the more evolved model discussed in Sect.~\ref{sec:results-evolution}. From an observational point-of-view, we seek answers to two questions: 1) ``Is \storm capable of predicting proper frequencies for low-order modes from its perturbative treatment of the rotational deformation of the star?'' and 2) ``When 1D asteroseismic modelling is performed and any residuals in the mode asymmetries are attributed to the presence of a magnetic field, how accurate would a magnetic field measurement then be?''. In Section~\ref{sec:num}, we discuss the numerical setup and work flow, in Section~\ref{sec:results-rotation} we discuss the differences between 1D and 2D methods for zero-age main sequence models and discuss the implications for measuring internal magnetic fields in Section~\ref{sec:results-magnetic}. Furthermore, in Section~\ref{sec:results-evolution} we discuss the effects of the nuclear evolution during the main sequence on our results, and finally, we conclude in Section~\ref{sec:conclusions}.

\section{Numerical setup} \label{sec:num}
We compute steady state models using the 2D stellar structure and evolution code \ester\footnote{\url{https://ester-project.github.io/ester/}}, \texttt{r23.09.1-evol} \citep{EspinosaLara2013, Rieutord2016, Mombarg2023-ester}. We limit ourselves to chemically homogeneous ZAMS models, which we compute for a 12\Msun star and rotation rates of 0, 5, 10, 15 and 20\percent of the critical Keplerian rotation frequency, $\Omega_{\rm c, Kep} = \sqrt{GM_\star/R_{\rm eq}^3}$. 
Our argument not to go beyond this rotation regime for the time being is twofold. From an astrophysical standpoint, the
population study of more than a hundred $\beta$\,Cephei pulsators carried out by \cite{Fritzewski2025} shows that their distribution of $v/v_{\rm c, Kep}$ peaks around 0.2. 
\conny{The practical reason is that we can no longer unambiguously identify the corresponding mode frequencies between the 1D and 2D pulsation calculations for rotation rates above $v/v_{\rm c, Kep}=0.2$.}
Table~\ref{tab:Omega} provides a summary of the deformation and rotation velocities. The mass of 12\Msun is chosen to be close to the asteroseismically inferred mass of the $\beta$~Cephei pulsator HD192575 \citep{Burssens2023, Vanlaer2025-modelling}. Mode multiplets have been identified in the pulsation spectrum of this star for modes of spherical degree $\ell = 1$ and 2. The rich pulsation spectrum of HD192575 makes it a prime target for modelling the interior rotation profile \citep{Vanlaer2025-modelling} and possibly measuring the internal magnetic field strength. 

\subsection{2D computations with TOP}
The \topc code can directly take an \ester model as input and uses the same 
spatially meshed grid, as the solvers of both codes are based on spectral methods. \edit{The \ester grid is based on a multi-domain approach to deal with strong variations and possible discontinuities (in the radial direction) in the solution. Within each domain, the stellar quantities are sampled over the Gauss-Lobatto collocation points associated with Chebyshev polynomials, with appropriate interface conditions between the domains \citep{Rieutord2016}. The first domain is placed over the convective core, and the boundaries of   the other domains are placed such that the pressure or temperature ratio between the inner and outer boundary remains roughly constant across domains.}

\begin{figure}
    \centering
    \includegraphics[width=\linewidth]{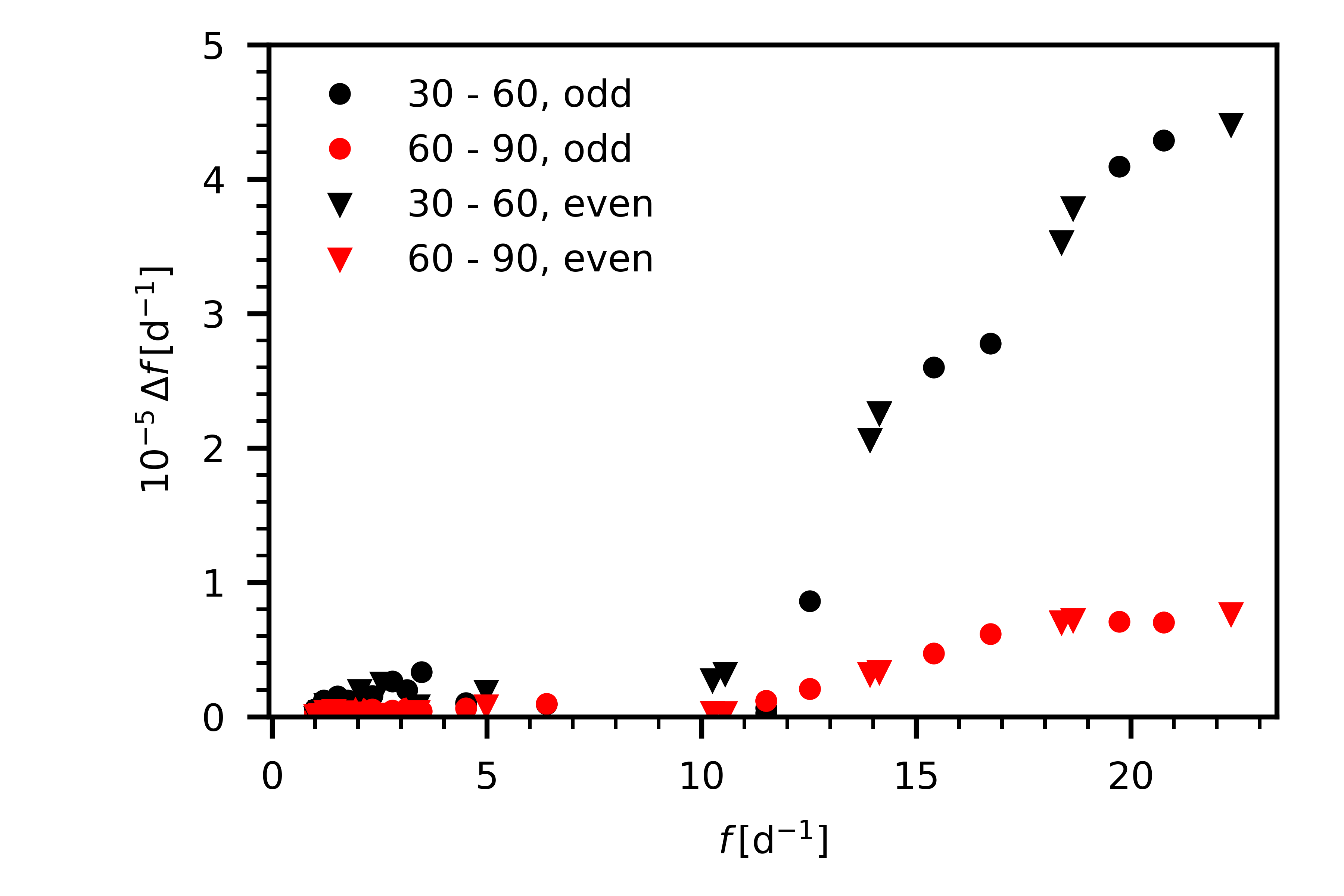}
    \caption{Frequency differences for different radial resolutions of the \ester/\topc grid. Frequency differences between 30 and 60 radial points per domain (in black), and differences between 60 and 90 points per domain (in red). }
    \label{fig:f-res}
\end{figure}

\begin{figure*}
    \centering
    \includegraphics[width=2\columnwidth]{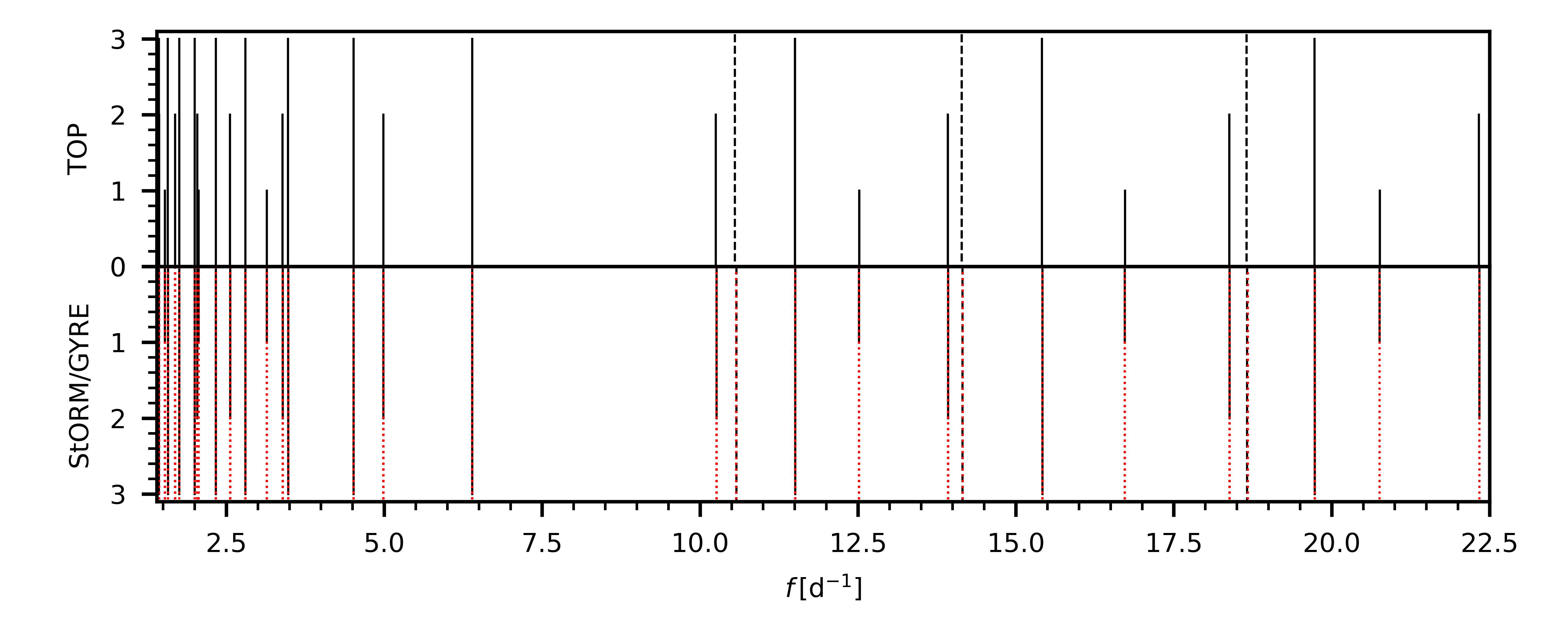}
    \caption{Frequency spectra of a non-rotating \ester model computed with \topc (top panel) and \storm (bottom) panel. The \gyre frequency spectrum is indicated with red dotted lines. The ordinate corresponds to the value of $\ell$. The radial modes are indicated with black dashed lines. }
    \label{fig:tvs}
\end{figure*}

We also couple the 1D pulsation code \gyre \citep{Townsend2013} to the non-rotating \ester model. For the \gyre computations, we assume a uniform rotation profile, where the traditional approximation of rotation (TAR) is used for the g~modes and the Coriolis force is neglected for the p~modes. It should be noted that \gyre (and \storm) assumes a mass distribution that is only dependent on the radius, which is not the case for rotating 2D models. In Fig.~\ref{fig:f-res}, we show the resulting frequency differences when the number of radial points of the \ester grid (same as the \topc grid) is increased. Fixing the number of domains -- bounded by isobars --  equal to 12, the differences are less than $10^{-5}\,{\rm d^{-1}}$ when the radial resolution is further increased from 60 points per domain to 90. For the 1D finite differences codes, the relevant physical quantities from \ester are interpolated to 2000 points in the radial direction. We only couple non-rotating \ester models to the 1D pulsation codes by generating a GYRE Stellar Model\footnote{\url{https://gyre.readthedocs.io/en/stable/ref-guide/stellar-models.html}} (GSM) from the \ester models using the interpolated quantities in an arbitrary $\theta$-direction. Also \storm is compatible with this GSM file format. In Fig.~\ref{fig:tvs}, we compare the frequency spectra of \topc and \storm (see Section~\ref{sec:storm}) for $\ell \in [0,1,2,3]$ computed from a non-rotating \ester model. We find excellent agreement between the output of the two codes, with frequency differences between $8 \cdot 10^{-6}$ and $0.0023\,{\rm d^{-1}}$. Similar frequency differences are found between \topc and \gyre (red dotted lines in the bottom panel of Fig.~\ref{fig:tvs}).

Rotation couples modes of different $\ell$ and $n_{\rm pg}$, and therefore, mode labelling in 2D is not trivial, making it challenging to identify modes belonging to the same multiplet as the mode spectrum is a priori infinitely dense in the gravito-inertial range of frequencies. Below, we describe our workflow to extract the mode splittings and their asymmetries.

\begin{itemize}
    \item Using \gyre, we scan a frequency interval such that all radial orders of interest are contained within the frequency interval. 
    \item Using \topc, we scan this frequency interval using a low angular resolution $n_\theta = 2$, where the highest spherical degree taken into account in the coupling is given by $\ell_{\rm max} = |m| + 2n_\theta$. \topc requires a frequency shift around which the code should find $n_{\rm sol}$ eigenmodes. We scan the interval of 1 to 6.5\cpd in steps of 0.5\cpd and 10 to 20\cpd in steps of 2\cpd, requesting 3 eigenmodes per frequency shift. The low-resolution mode enables us to identify the locations of the large-scale modes that we are interested in, while saving CPU time. 
    \item Since the frequencies and eigenvectors of these large-scale modes are not fully resolved with this resolution, we scan again around the found frequencies, but with a higher resolution $n_\theta = 10$. With this higher resolution, however, also small-scale modes are found close to the large-scale mode. 
    \item In order to filter out these small-scale modes from the large-scale ones, we assign a `pseudo' $\hat{\ell}$ and $\tilde{n}_{\rm pg}$, by simply counting the number of nodes in the radial and angular directions. It should be noted that this way of mode labelling is more ad-hoc compared to that used by \cite{Mirouh2019} to label island modes, as the number of nodes may depend on the radius where we do the counting. Nevertheless, we find that in most cases $\hat{\ell} = n_{\rm nodes, \theta} + 2|m|$ allows us to identify the multiplets, where $n_{\rm nodes, \theta}$ is the number of nodes in the $\theta$-direction over both hemispheres\footnote{Where possible, manual mode identification was done for a few cases with unsuccessful frequency correspondence by visually inspecting the eigenfunction of candidate modes. }. We then filter out all modes $\hat{\ell} >2\ell$ and $\hat{n}_{\rm pg} \geq 15$. 
    \item Lastly, we match the remaining modes with the labelled ones from \gyre by simply selecting the mode with the same $m$ that has a frequency difference below a certain threshold. Now that we have identified the multiplets in the \topc spectrum, we can compute the mode asymmetries.
\end{itemize}
The modes computed by \topc are classified as either `even', symmetric around the equator, or `odd', anti-asymmetric around the equator. In the case of $\ell = 1$, the $m=0$ modes are odd, and the $m=\pm 1$ modes are even. Likewise, for $\ell = 2$, the $m=0$ and $m=\pm2$ modes are even, and the $m=\pm 1$ modes are odd. 

\subsection{1D computations with StORM} \label{sec:storm}
We also compute the frequencies with the newly developed 1D \storm code\footnote{\url{https://stellar-oscillations.org/}} \citep{Vanlaer2025-storm}. \storm solves the adiabatic oscillation equations, including the effects of rotation. It includes the Coriolis acceleration and a perturbative approximation for the centrifugal deformation of the star. This is accomplished in two steps. First, the oscillation equations are solved directly for a single spherical degree, only including the Coriolis acceleration terms. Second, the solutions from the simplified oscillation equations are perturbed taking the stellar deformation into account, as well as the  coupling between different spherical degrees due to the contribution of the Coriolis acceleration and the toroidal components \citep[e.g.][]{saio1981,LeeBaraffe1995}. Since \storm uses a 1D stellar model as input, the stellar deformation is approximated using the Chandrasekhar-Milne expansion up to the $P_2$ term \citep{chandrasekhar_equilibrium_1933}. It should be noted that the 1D computations with \storm are much faster than the 2D computations with \topc (order seconds compared to $\sim$\,30\,min total computation time on 4 CPUs for our case when scanning the entire frequency range for even or odd modes).

\begin{table}[]
    \centering
        \caption{Ratio radius at the pole to radius at equator for \ester models with different rotation rates.}
    \begin{tabular}{c|c|c|c}
    \hline \hline
        $\Omega/\Omega_{\rm c, Kep}$ & $R_{\rm e}/R_{\rm p}$ & $\Omega~[{\rm d}^{-1}]$ & $v_{\rm eq}~[{\rm km\,s^{-1}}]$ \\
        \hline
        0.05 & 1.0009 & 0.18 & 37 \\
        0.10 & 1.0052 & 0.36 & 74 \\
        0.15 & 1.0111 & 0.53 & 111 \\
        0.20 & 1.0196 & 0.70 & 148 \\
        \hline
    \end{tabular}

    \label{tab:Omega}
\end{table}

\section{Mode asymmetries} \label{sec:results-rotation}

In this section, we compare the predicted mode asymmetries of \storm and \topc. We compute mode asymmetries for $\ell \in [1, 2]$ and $n_{\rm pg} \in [-5,-4,-3,-2,-1,0,1,2]$, corresponding to typically observed values in $\beta$~Cephei pulsators \citep{StankovHandler2005,Fritzewski2025}. The computations of \storm were done for a non-rotating \ester model, where we consider the rotational effect only in the pulsation computations. We do so by taking
a uniform rotation profile with equal fraction of the critical rotation frequency compared to the rotating \ester model. The computations with \topc use the differential rotation profile that follows from the baroclinic torque as computed by \ester \citep{EspinosaLara2013}. 

\begin{figure}[h]
    \centering
        \includegraphics[width=\linewidth]{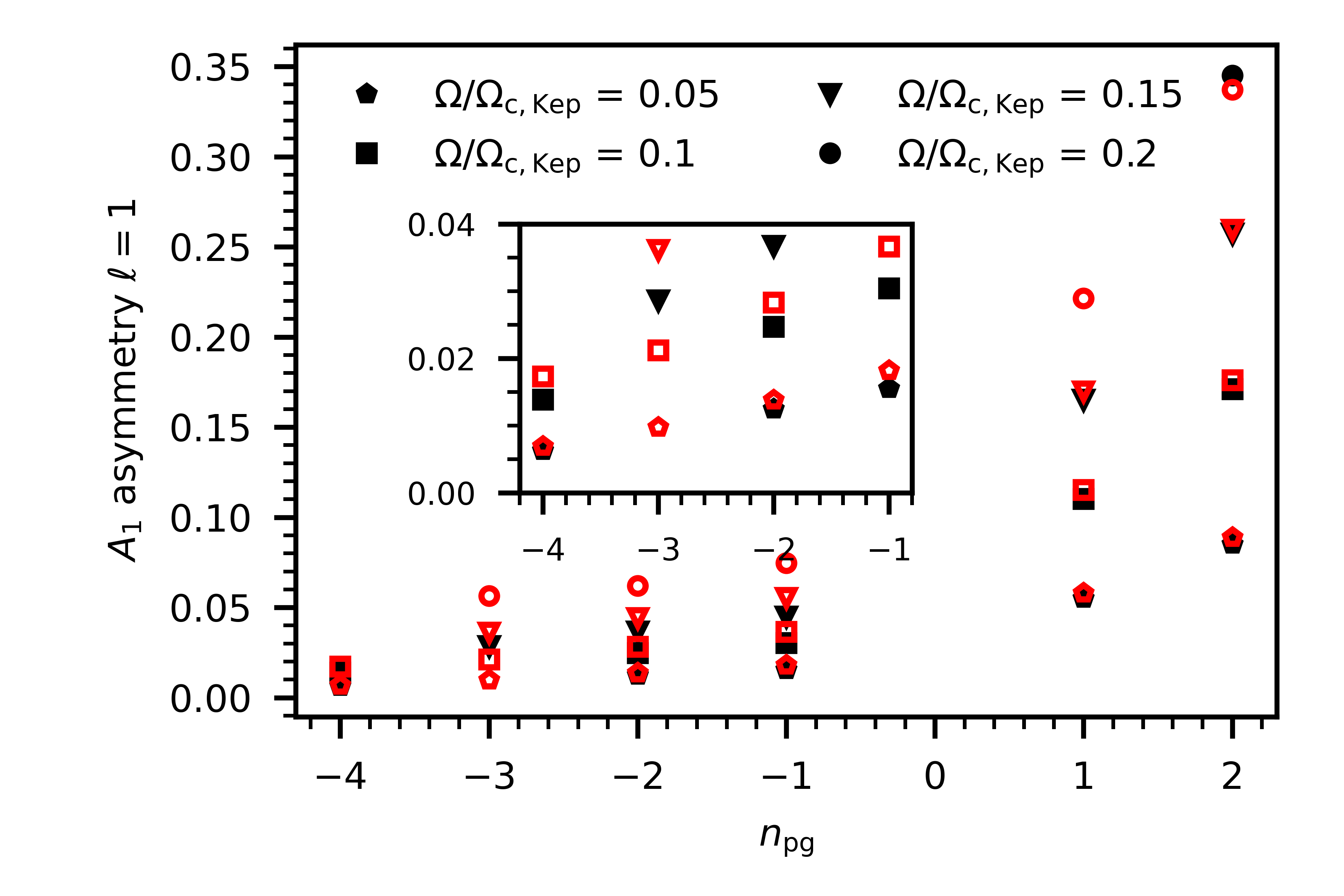}

    \caption{Predicted asymmetry parameters (dimensionless) for $\ell = 1$ computed with \topc (in black) and \storm (in red) as a function of radial order. }
    \label{fig:A_odd}
\end{figure}

\begin{figure}[h]
    \centering
    \begin{subfigure}{0.49\textwidth}
        \centering
        \includegraphics[width=\linewidth]{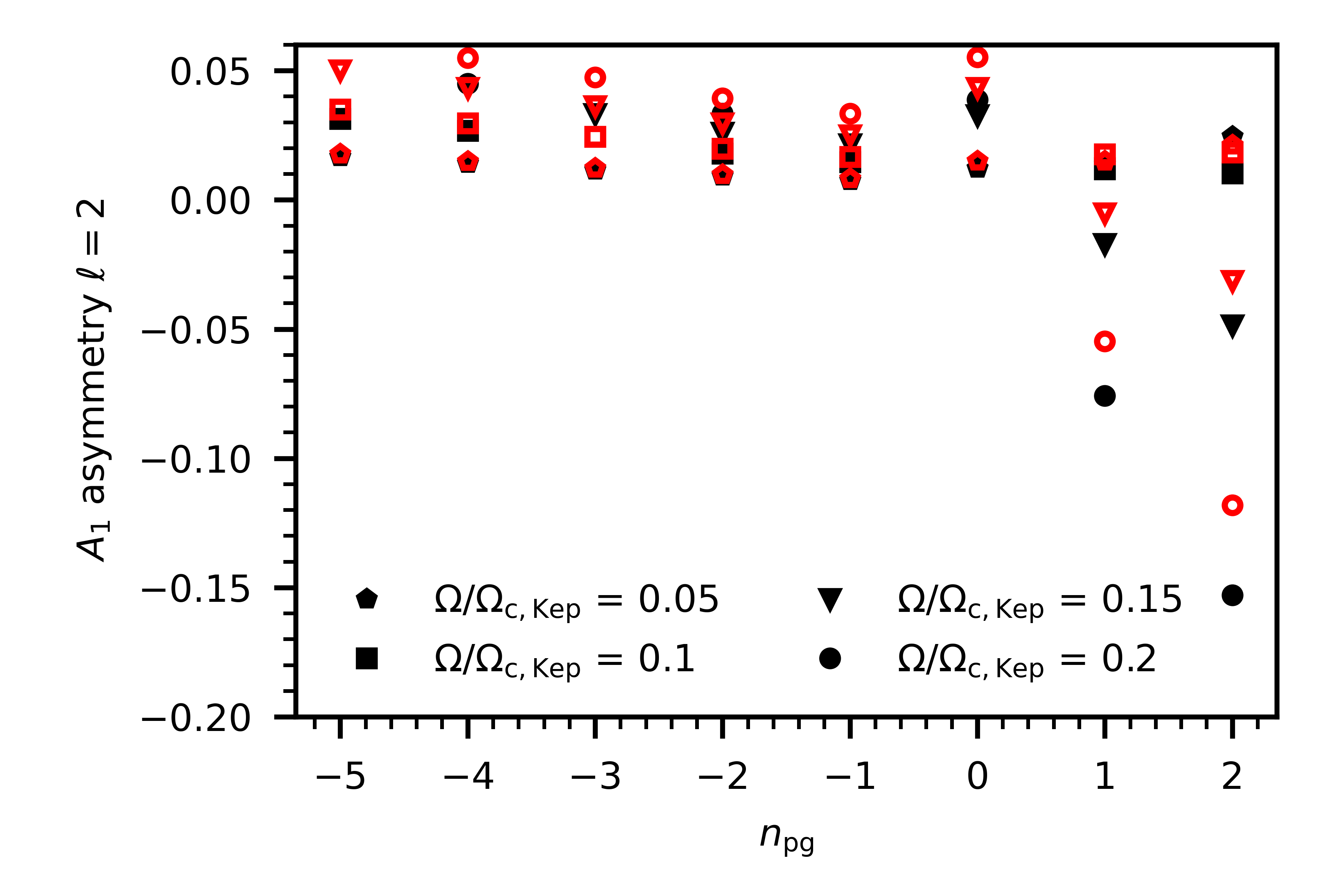}
        \label{fig:sub1}
    \end{subfigure}
    \hfill
    \begin{subfigure}{0.49\textwidth}
        \centering
        \includegraphics[width=\linewidth]{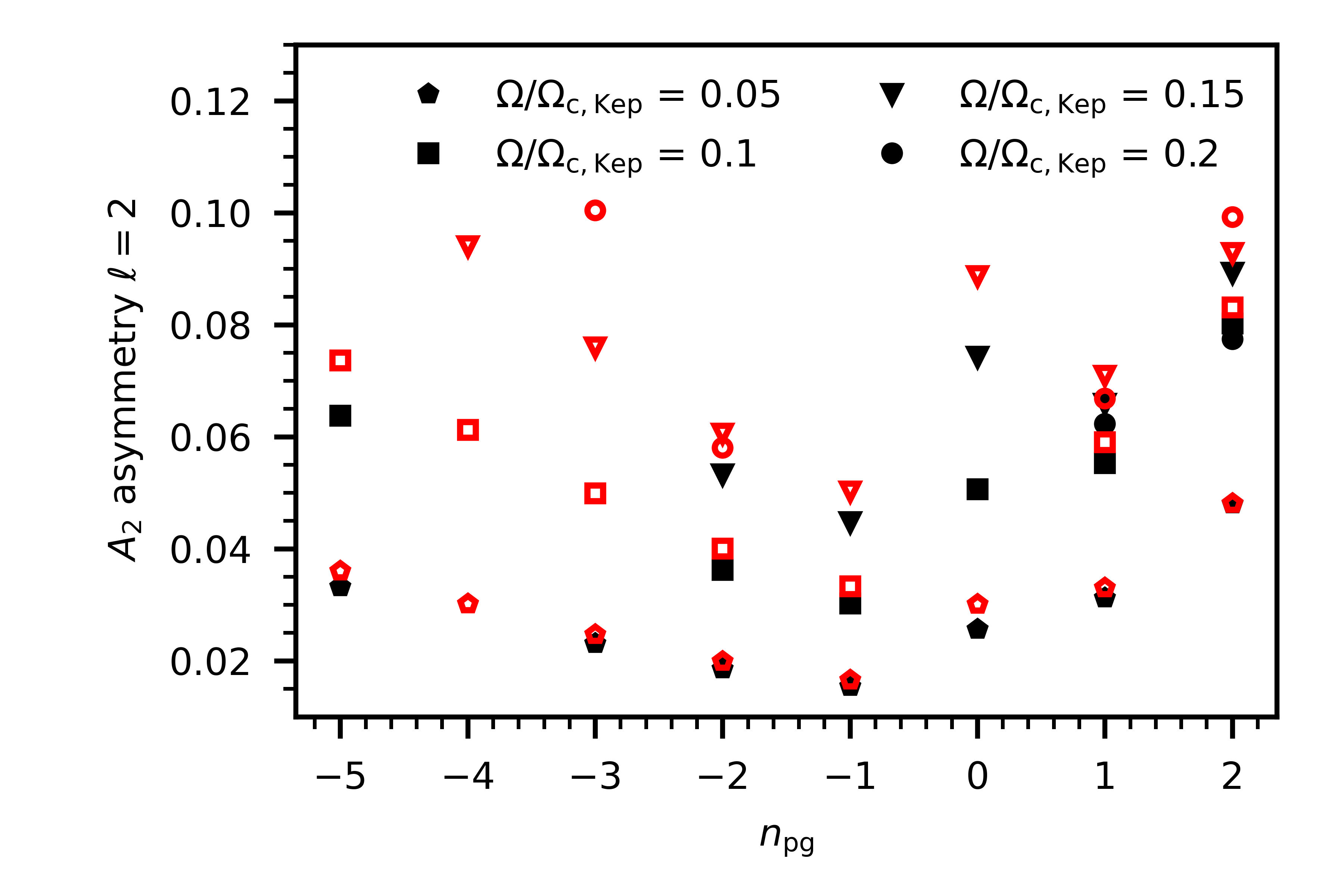}
        \label{fig:sub2}
    \end{subfigure}
    
    \caption{Predicted asymmetry parameters for $\ell = 2$ computed with \topc (in black) and \storm (in red) as a function of radial order. }
    \label{fig:A_even}
\end{figure}

We define the relative asymmetry parameter as follows,
\begin{equation} \label{eq:A}
    A_{|m|} = \frac{2 f_{0} - f_{-m} - f_{+m}}{f_{+m} - f_{-m}},
\end{equation}
where $f_{-m} < f_0 < f_{+m}$ ($m > 0$ for prograde modes).
The results are shown in Figs.~\ref{fig:A_odd} and \ref{fig:A_even}. Overall the predictions between the 1D perturbative method and full 2D are in good agreement. The largest differences are observed for the p modes and increase with increasing rotation. We confirm that the asymmetries of the low-order $\ell =1$ modes are strictly positive, while negative asymmetries $A_1$ occur for the $\ell = 2$ p modes for rotation rates above 10\percent\ of the critical rotation. In Fig.~\ref{fig:A_even}, the additional $A_{2}$ asymmetry parameters are shown, which are also strictly positive. In some cases, particularly for the model with the highest rotation rate, some frequencies of a multiplet could not be clearly identified in the \topc frequency spectrum as a result of additional modes with similar frequency and a significant change in the eigenfunction compared to the eigenfunction for zero rotation.  

\begin{figure}[h]
    \centering
    \begin{subfigure}{0.49\textwidth}
        \centering
        \includegraphics[width=\linewidth]{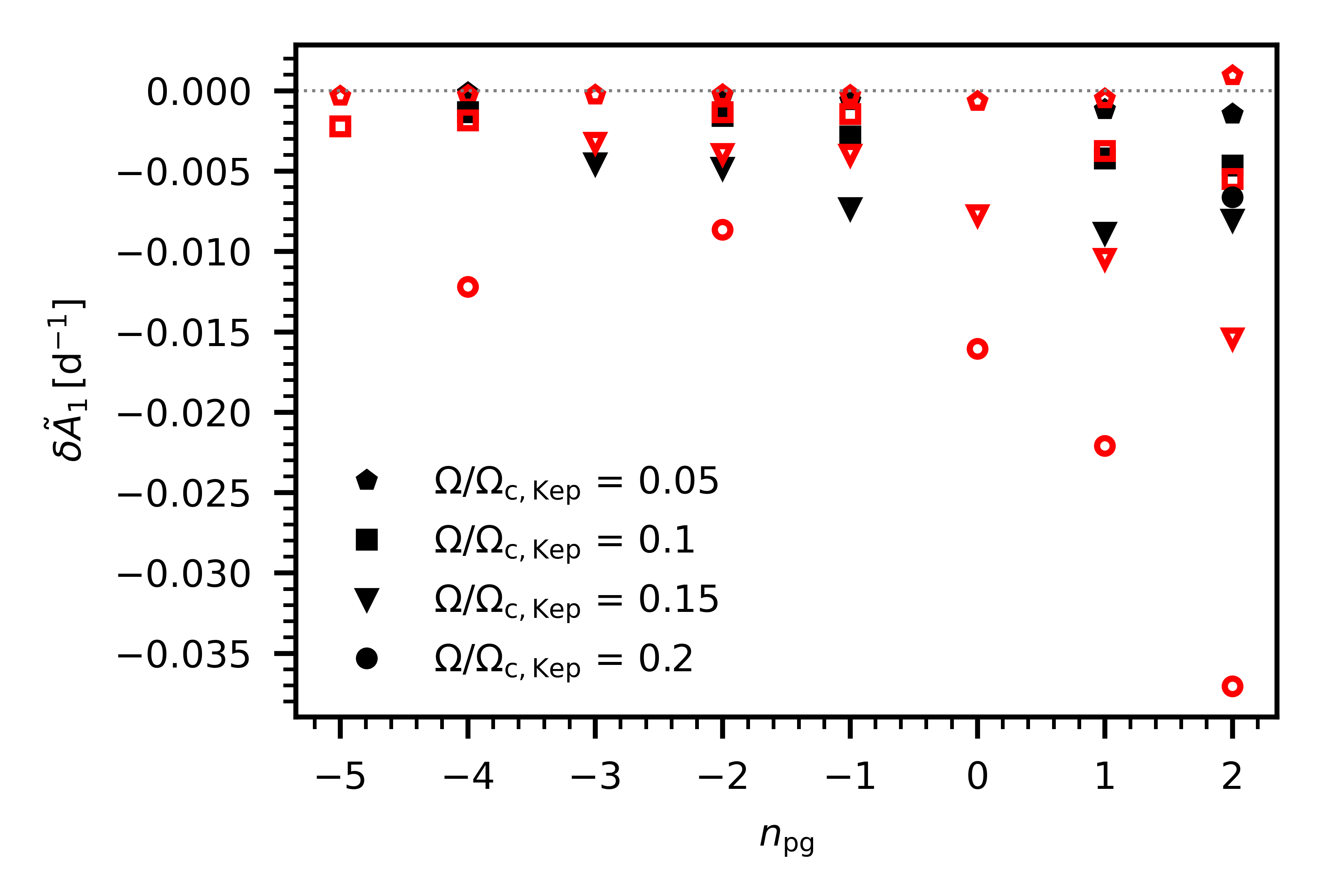}
        \label{fig:sub1}
    \end{subfigure}
    \hfill
    \begin{subfigure}{0.49\textwidth}
        \centering
        \includegraphics[width=\linewidth]{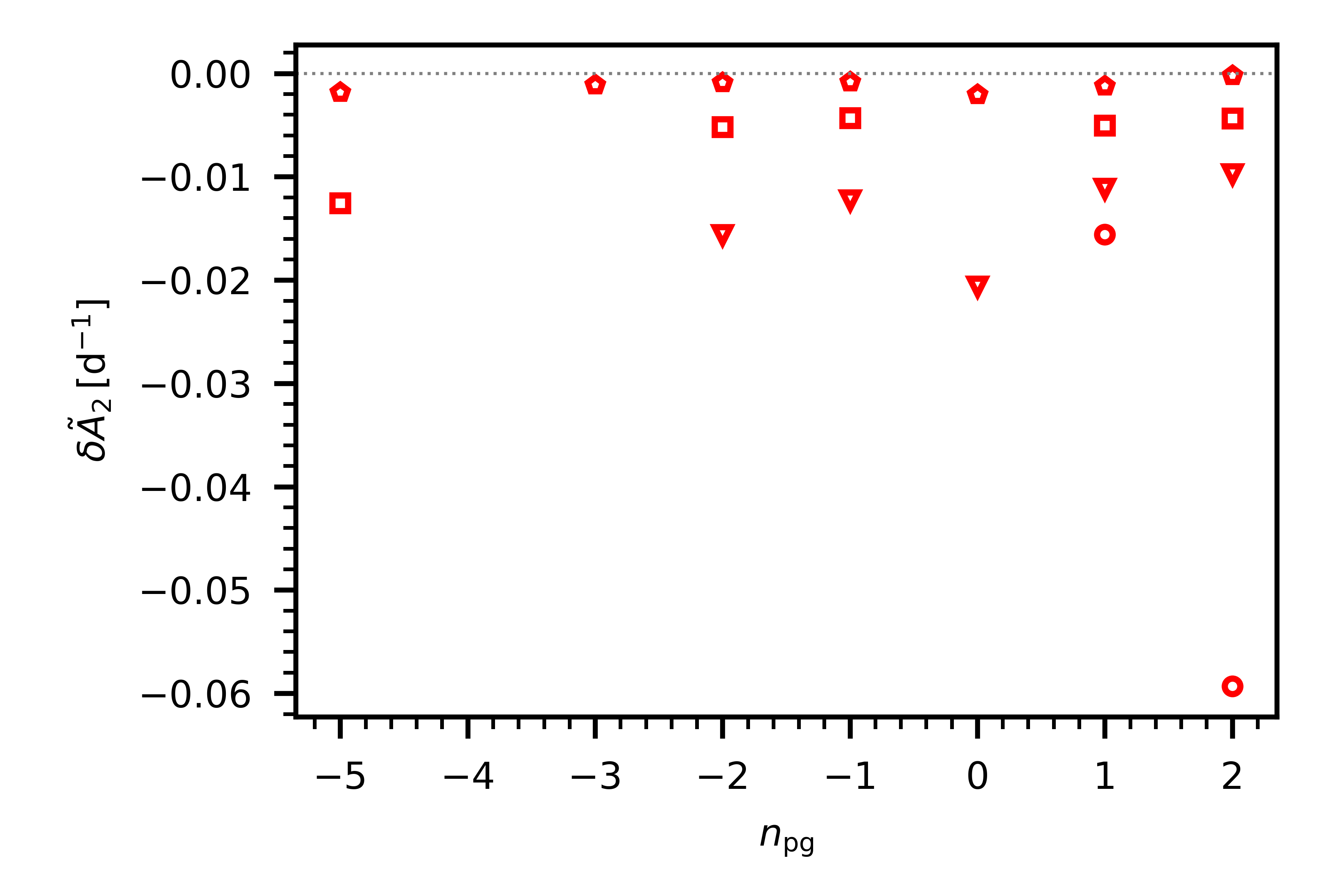}
        \label{fig:sub2}
    \end{subfigure}
    
    \caption{Differences in the predicted asymmetry (\topc $-$ \storm) as a function of radial order. Black symbols corresponds to $\ell =1$, red symbols to $\ell = 2$.}
    \label{fig:deltaA}
\end{figure}

The differences in the asymmetries\footnote{We use a tilde to distinguish between the dimensionless asymmetry ($A$) and the asymmetry in units of frequency ($\tilde{A}$). } in units of ${\rm d^{-1}}$ (numerator in Eq.~(\ref{eq:A})) are shown in Fig.~\ref{fig:deltaA}. We find that differences in the asymmetry are mostly limited to $0.06\,{\rm d^{-1}}$, and below $0.04\,{\rm d^{-1}}$ for rotation rates $\le 15 \%$ of the critical rotation. 
\conny{This is within the realm of theoretical uncertainties on predictions of mode frequencies due to limited knowledge on the input physics of the models \citep{Aerts2018}. However,  these uncertainties will be very similar for all the frequencies in Eq.\,(\ref{eq:A}) and thus will not hamper interpretations of asymmetries in observed mode multiplets of real stars.  In absolute terms, these differences in asymmetries between \topc and \storm predictions are smaller than the actual measured rotation frequency of $\beta$~Cephei pulsators such that identifications of $m-$values is also not an issue in practice, except perhaps for the slowest rotators such as HD\,129929 \citep{Aerts2003,Dupret2004-HD129929}.}
In most cases, the difference $\delta \tilde{A}$ is positive, meaning the 1D method in general tends to underestimate the asymmetries. 

We estimate the contribution of the radial differential rotation (2D method) versus uniform rotation (1D setup) to the total differences in frequencies by evaluating $\alpha = |r \partial_r\Omega|/\bar{\Omega}$, where $\bar{\Omega}$ is the average rotation frequency. For our models used in this work, $\alpha < 0.1$, meaning that we do not expect that the difference in the rotation profile between the 1D and 2D method has a very large effect.

\section{Effect on measurements of magnetic fields} \label{sec:results-magnetic}

In this Section, we quantify what the differences in mode asymmetries predicted from 1D or 2D methods could imply for the detection and measurement of an internal magnetic field. If an internal magnetic field is invoked to explain any residuals in observed asymmetries after correcting for the rotation, what would be the difference in the measured magnetic field characteristics originating from the differences between the 1D and 2D frequency predictions adopted for the rotational asymmetry part?

The computation of the magnetic asymmetries are computed with the \texttt{magsplitpy} code, and we refer the reader to \cite{Das2024} for the details of the code. The code uses the same description for the magnetic field topology as outlined in \cite{prat2019} and \cite{bugnet2021}. The chosen field configuration, initially derived in \cite{DuezMathis2010}, is composed of poloidal and toroidal components such that the magnetic field is stable over evolutionary timescales. However, unlike in \cite{bugnet2021}, where the magnetic field beyond the radiative core is zero for their application to red giant stars, we extend the field all the way through the envelope (vanishing at the stellar surface), as the modes studied here are sensitive to larger portion of the star. The magnetic asymmetries are computed for different inclinations of the field with respect to the star's rotation axis, referred to as the obliquity angle, $\beta$, which is known to crucially modulate the magnetic asymmetries as shown in \citet{Loi2021, li2022, MathisBugnet2023, Das2024}.

\begin{figure}
    \centering
    \includegraphics[width=\linewidth]{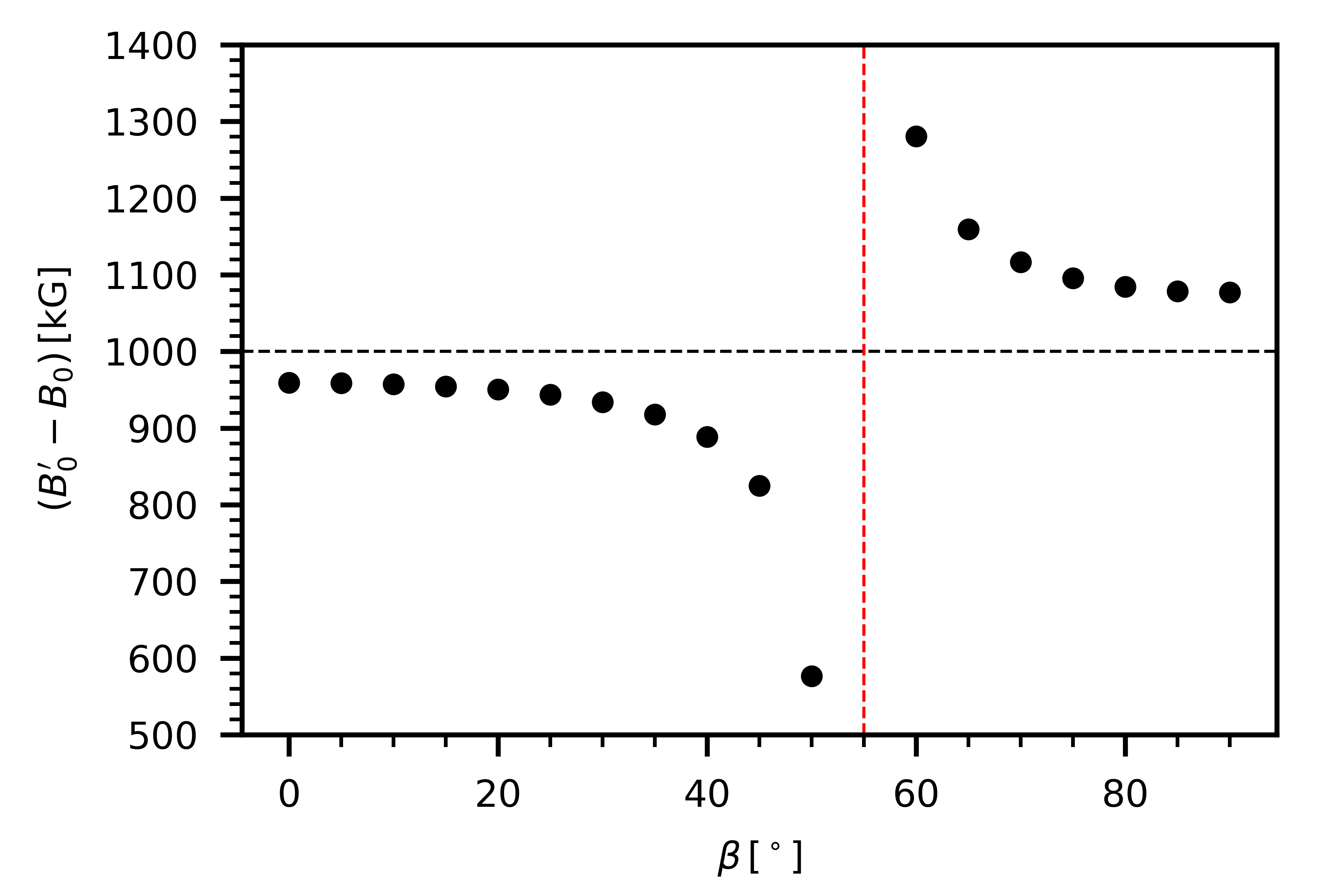}
    \caption{Difference between the measured magnetic field strength ($B_0^\prime$) and the actual one ($B_0 = 1\,{\rm MG}$) as function of the obliquity angle $\beta$. Here, $A_1$ asymmetries were used for $n_{\rm pg} = 2$, and $\delta \tilde{A}$ is based on a model rotating at 5\percent\ of the critical rotation frequency. The critical angle where the magnetic asymmetry changes sign is indicated with a red vertical line. }
    \label{fig:beta}
\end{figure}

The magnetic asymmetries scale with amplitude of the magnetic field squared, $\tilde{A} \propto B_0^2$ (for the assumed topology of the magnetic field). Therefore, for a given multiplet and field obliquity angle $\beta$, the amplitude of the magnetic field can be over- or underestimated if the calculated asymmetry resulting from rotation is inaccurate. For example, a difference $\delta \tilde{A}$ in the asymmetry predicted between 1D and 2D methods will result in a difference of the measured magnetic field,
\begin{equation}
    B_0^\prime  = B_0 \sqrt{\frac{\tilde{A}_{\rm 1D} + \delta \tilde{A}}{\tilde{A}_{\rm 1D} }},
\end{equation}
where $B_0^\prime - B_0$ is the error\footnote{The terms `error' and `uncertainty' mentioned here refer to over- or underestimation of the magnetic field strength between 1D and 2D methods for the rotation. This should not be confused with an observational measurement precision.} on the inferred field strength when relying on 1D perturbative methods to treat the rotational effects. From an observational point-of-view, $\tilde{A}_{\rm 1D}$ represents the residual asymmetry after accounting for the rotation, which is then assumed to be the result of magnetic effects. Hence, when $\delta \tilde{A} > 0$, measurements of the field strength will be overestimated. As the obliquity angle increases, the magnetic asymmetries for a dipolar field will decrease until a critical angle $\sim 55^{\circ}$ where there is a change of sign \citep{li2022, MathisBugnet2023, Das2024}. Therefore,  accurate measurement of $B_0$ when $\beta$ is close to this critical value is impossible, as is shown in Fig.~\ref{fig:beta}. In the example shown in this figure, $B_0 = 1\,{\rm MG}$, which gives large enough $\tilde{A}_{\rm 1D}$ to prevent the contribution of $\delta \tilde{A}$ to flip the sign. 

\begin{figure}
    \centering
    \includegraphics[width=\linewidth]{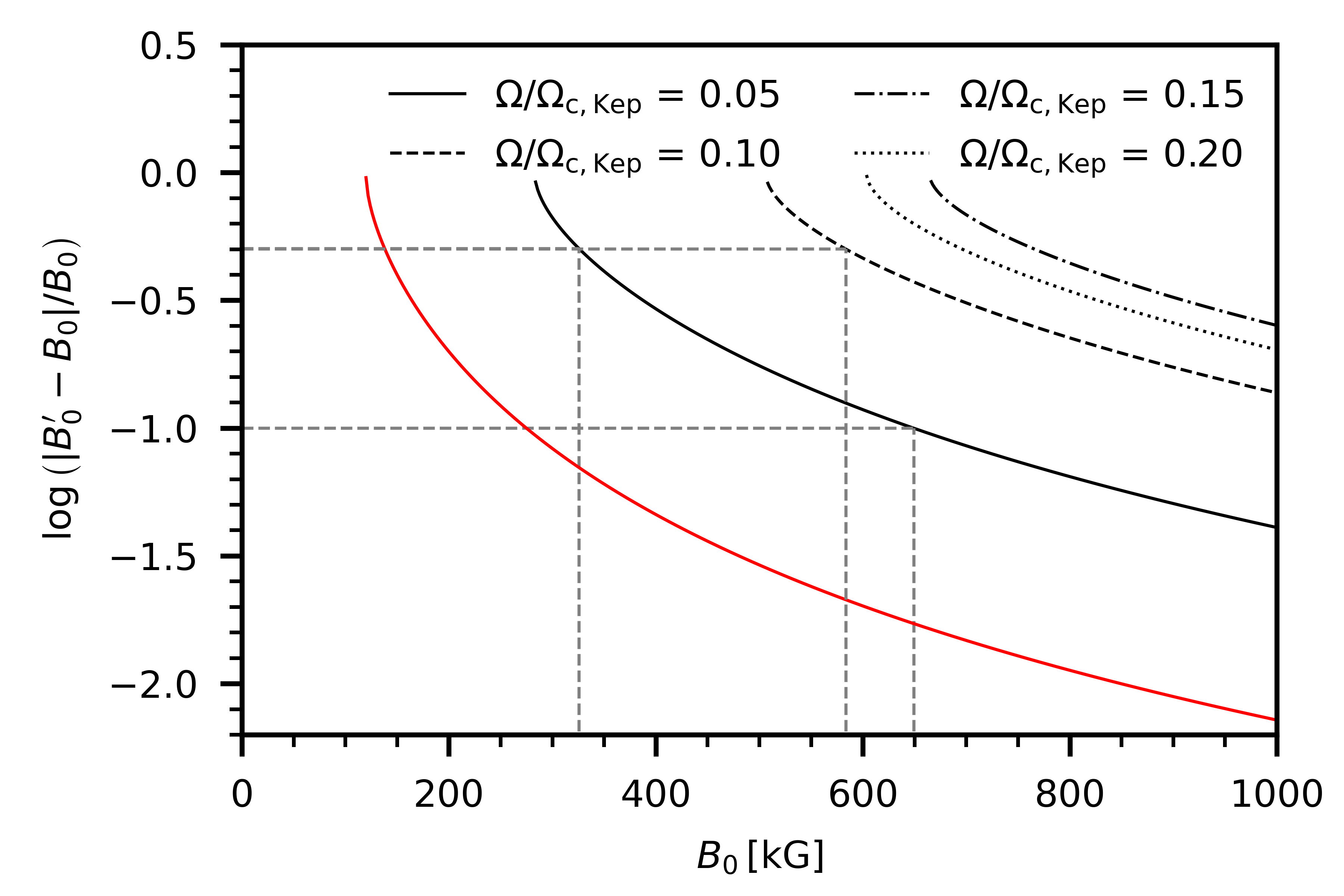}
    \caption{Error of the measured magnetic field strength as a function of the actual magnetic strength. The black curves correspond to $A_1$ asymmetries ($\ell =1$), the red curve to $A_2$ asymmetries. Field strengths fractional errors corresponding to 50\% and 10\% are indicated with gray dashed lines.}
    \label{fig:deltaB0}
\end{figure}

We now quantify the error on the inferred $B_0$ when $\beta = 0^{\circ}$ for each of the radial orders for which we could determine $\delta \tilde{A}$. The magnetic asymmetries were computed with \texttt{magsplitpy} for $B_0 = 75\,{\rm kG}$. We use $\tilde{A}_{\rm 1D}(B_0) = (B_0/75\,{\rm kG})^2 \tilde{A}_{\rm 1D}(75\,{\rm kG})$ to scale the asymmetries to any value of the magnetic field strength. 
Figure~\ref{fig:deltaB0} shows the error of the measured magnetic field strength as a function of the actual one for the most optimistic case, namely $n_{\rm pg} = 2$. In this case, field strengths above $\sim 300\,{\rm kG}$ could be measured with an error better than 50\percent, and field strengths above $\sim 700\,{\rm kG}$ could be measured with an error better than 10\percent. For comparison, measured surface magnetic field strengths for O-type are typically on the order of a few kG, and $\sim$4-10\,kG for early B-type stars \citep{Shultz2019}. We note that this minimum $B_0$ scales inversely with the fraction of critical rotation. The range of $B_0$ covered in Fig.~\ref{fig:deltaB0} is simply chosen to cover a large range and the higher values may be higher than what is typical for real stars. The lower end points of the curves indicate where the square-root term in Eq.~(\ref{eq:A}) becomes negative. For the other radial orders $n_{\rm pg} \in [-5, ..., 1]$, the predicted magnetic asymmetries of this \ester ZAMS model are so small that even at $0.05\,\Omega/\Omega_{{\rm c, Kep}}$ the uncertainty on the rotational asymmetry prevents any magnetic field measurement below $1\,{\rm MG}$ with reasonable uncertainty. Therefore, even though the difference in the predicted rotational asymmetry between 1D and 2D methods is larger in low-order p~modes compared to low-order g~modes, the magnetic asymmetries are also larger and scale in such a way that modes with $n_{\rm pg} \ge 2$ provide the best diagnostic for potential measurements of internal magnetic fields in $\beta$~Cephei stars. 

\section{Effect of nuclear evolution on the asymmetries} \label{sec:results-evolution}
We now aim to investigate how dependent the results presented in the previous sections are on the effect of the nuclear evolution during the main sequence. Achieving nuclear evolution with \ester with sufficient mesh resolution for the pulsation computations is numerically challenging and time consuming\footnote{In order to reduce the number of steps needed to reach convergence we relax the default tolerance with a factor 5.}. We have evolved a 12\Msun star with \ester up to a central hydrogen mass fraction of $X_{\rm c} \approx 0.4$, once for a non-rotating star and once for an initial rotation rate $0.1\,\Omega/\Omega_{\rm c, Kep}$ (at the ZAMS). Such an evolutionary stage ($X_{\rm c}$) is representative for the sample of $\beta$~Cephei pulsators of \citet{Fritzewski2025}. The \texttt{r23.09.1-evol} release of \ester that we are using here treats chemical mixing in the radiative envelope based on the work of \citet{Zahn1992} with a prescribed vertical eddy-viscosity associated with the vertical shear, whereas for the non-rotating case a constant diffusion coefficient is used that we here set equal to $10^4\,{\rm cm^2\,s^{-1}}$, based on measurements of \cite{Pedersen2021}. Therefore, there are some small differences in the stellar structure between the rotating and non-rotating model, but this should have a minimal effect on the mode asymmetries as the two models are still very similar in structure (fractional convective core mass 0.2532 and 0.2501, respectively). 

\begin{figure}[h]
    \centering
    \begin{subfigure}{0.49\textwidth}
        \centering
        \includegraphics[width=\linewidth]{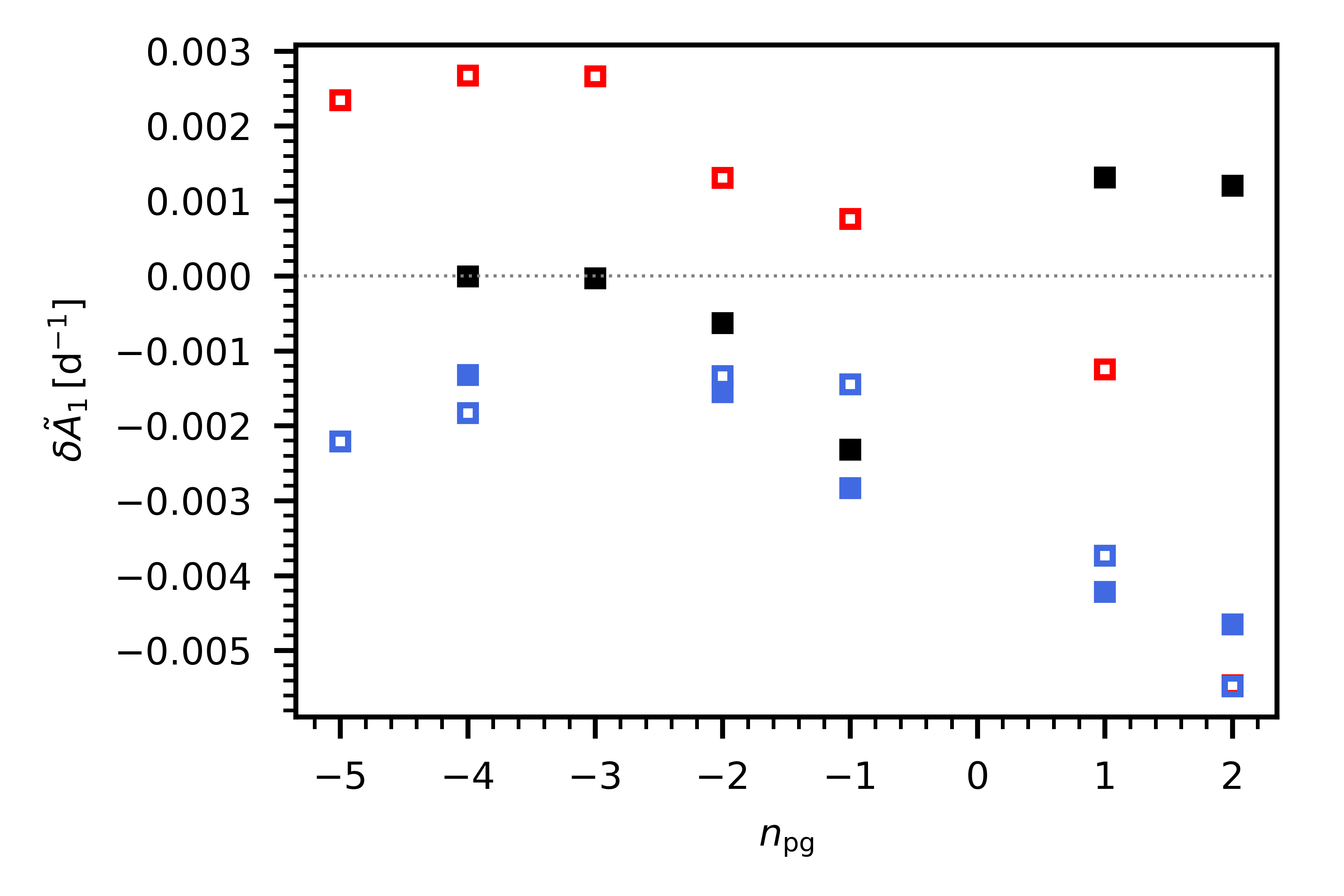}
        \label{fig:sub1}
    \end{subfigure}
    \hfill
    \begin{subfigure}{0.49\textwidth}
        \centering
        \includegraphics[width=\linewidth]{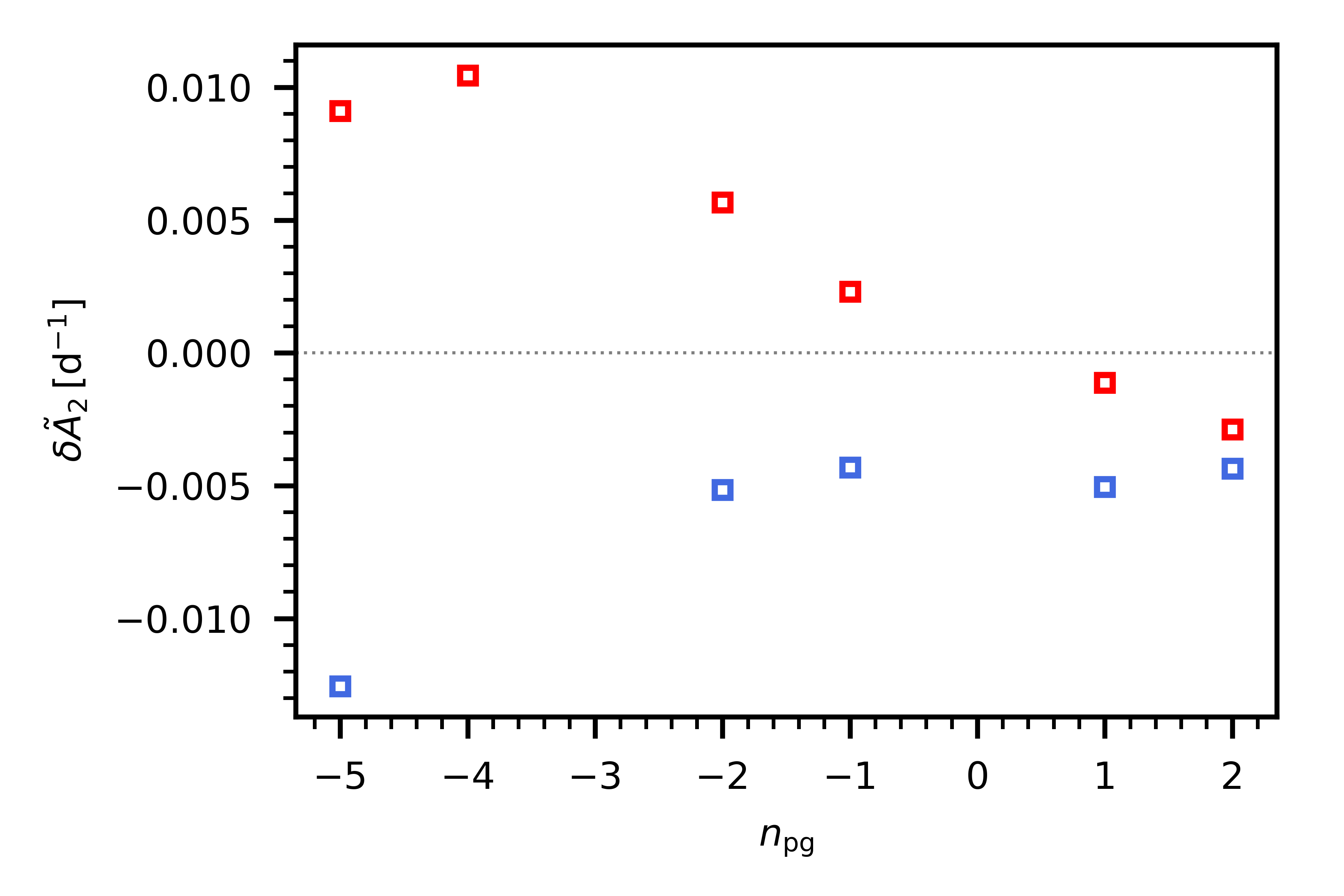}
        \label{fig:sub2}
    \end{subfigure}
    
    \caption{Same as Fig.~\ref{fig:deltaA}, but for a model with $X_{\rm c} \approx 0.4$. The fraction of critical rotation of this model is $0.0942\,\Omega_{\rm c, Kep}$. Black symbols corresponds to $\ell =1$, red symbols to $\ell = 2$. For comparison, the results from the ZAMS model at $0.1\,\Omega_{\rm c, Kep}$ from Fig.~\ref{fig:deltaA} are also shown in blue. }
    \label{fig:deltaA_mams}
\end{figure}

\begin{figure}
    \centering
    \includegraphics[width=\linewidth]{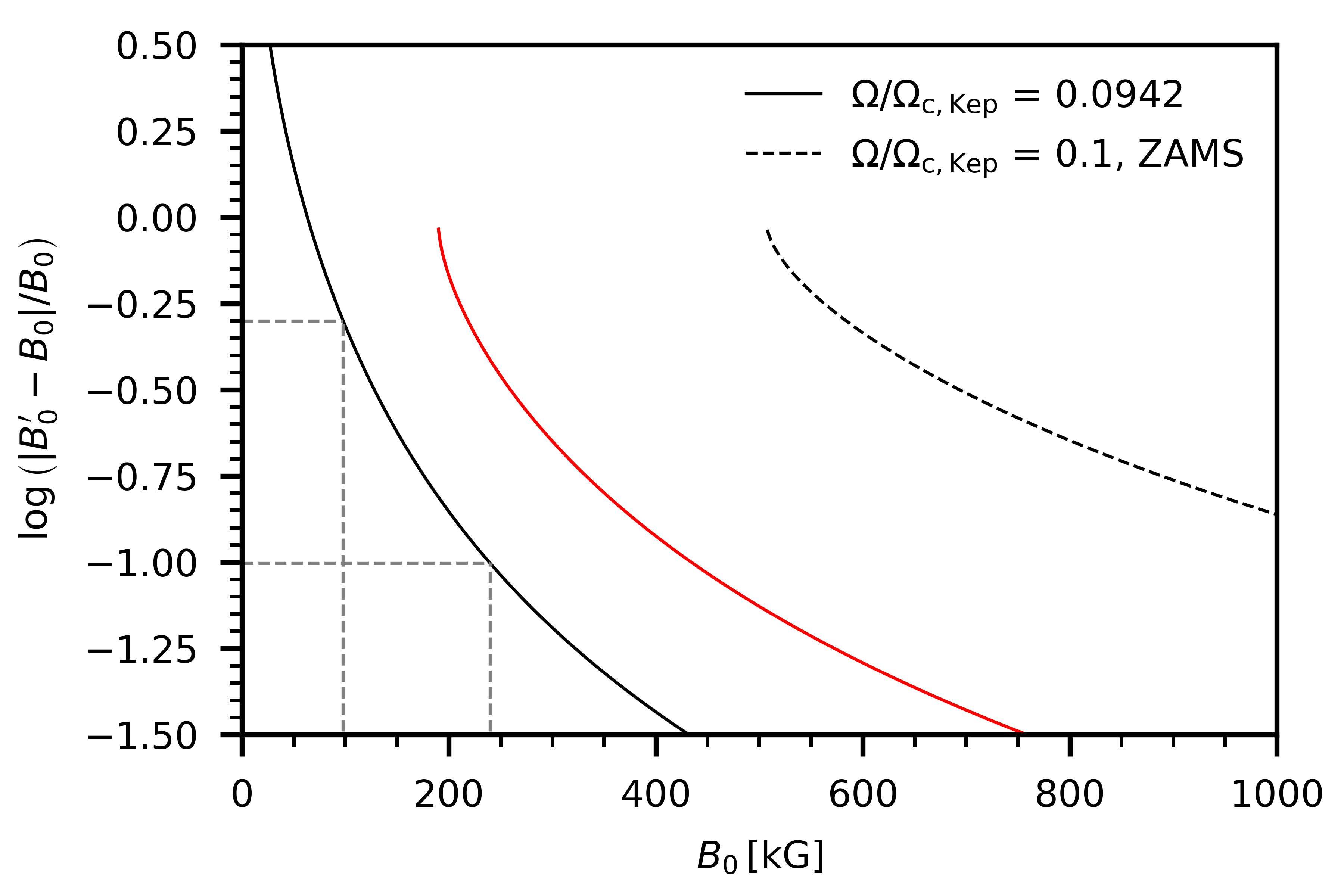}
    \caption{Same as Fig.~\ref{fig:deltaB0}, but for a model at $X_{\rm c} = 0.4$. For comparison, the results for the ZAMS model with a rotation rate of $0.1\,\Omega/\Omega_{\rm c, Kep}$ is shown again here. }
    \label{fig:deltaB0_MAMS}
\end{figure}

Figure~\ref{fig:deltaA_mams} shows the differences in the predicted asymmetries between 1D and 2D methods. While the model starts with a rotation rate of $0.1\,\Omega_{\rm c, Kep}$ at the ZAMS, the fraction of critical rotation decreases slightly, as is the case for \ester models above $\sim$8\Msun \citep{Mombarg2024-ester}. The evolved model that we take here rotates at $0.0942\,\Omega_{\rm c, Kep}$. For comparison, the differences in asymmetries for the ZAMS model rotating at $0.1\,\Omega_{\rm c, Kep}$ are also shown. As can be seen from Fig.~\ref{fig:deltaA_mams}, similar differences are found for the more evolved model compared to the ZAMS model.

Again, we investigate the implications for measurements of $B_0$. The magnetic splittings and asymmetries depend on the product of the mode kernel and the Lorentz stress tensor $\bold{B} \bold{B}$. We find that the effect of the evolution of the stellar structure plays a bigger role for the magnetic asymmetries compared to the rotational ones. The magnetic asymmetries of the more evolved model that we study here are about 4 to 18 times larger than those of the ZAMS model. In Fig.~\ref{fig:deltaB0_MAMS}, we show the error of the measured $B_0$ similar to Fig.~\ref{fig:deltaB0}. As can be seen, a magnetic field could be measured with a smaller error for the more evolved model, since the magnetic asymmetries become larger while the rotational ones remain similar. For the model rotating at $0.0942\,\Omega_{\rm c, Kep}$, 50\percent error could be achieved around 250\,kG instead of 600\,kG for the ZAMS model. Furthermore, we note that while for the ZAMS model the $A_2$ ($\ell = 2$) asymmetries would be more accurate compared to the $A_1$ asymmetries, it is the other way around
for the more evolved model.

\section{Conclusion} \label{sec:conclusions}
In this paper, we have compared a 1D perturbative method to compute eigenfrequencies of rotating stars with a non-perturbative 2D method (including 2D stellar equilibrium models). We presented predicted mode asymmetries of the rotational splittings for 12\Msun ZAMS models, covering low radial-order modes that are typically observed in $\beta$~Cephei pulsators \citep[e.g.][]{StankovHandler2005,Fritzewski2025}. We computed asymmetries for rotation rates up to 20\percent of the critical Keplerian rotation rate, above which it becomes difficult to identify multiplets in the predicted mode spectra of the g~modes. 
We show that the sign of the asymmetries is consistent between the 1D and 2D methods, and, therefore, stars that show asymmetries of an opposite sign could indicate the presence of a magnetic field. 

We have shown that when residual observed asymmetries -- after accounting for rotational effects with 1D methods -- are assumed to come from magnetic effects, the resulting measured magnetic field strength can change by orders of magnitude between 1D and 2D methods used for the rotation, even when the centrifugal deformation is minimal. However, we find that reasonably accurate measurements of the magnetic field strength are in principle possible for $n_{\rm pg} = 2$ (and likely higher radial orders, although unlikely to be excited) when the fraction of critical rotation is not more than 10\percent and the magnetic field strength is large enough (at least $\gtrsim 300\,{\rm kG}$).

As shown in this work, accurate measurements of an internal magnetic field relying on 1D perturbative methods to treat the rotation is extremely limited. Nevertheless, this of course does not imply that such methods have no scientific use. The \storm code that we used here is able to reproduce rotational asymmetries with an adequate precision in a fast manner. This makes it a valuable tool for asteroseismic measurements of rotational properties, with the capacity to treat large samples of stars. \storm's 1D perturbative approach for the centrifugal deformation provides appreciably better accuracy over current state-of-the-art pulsation codes that do not account for this deformation. We do note, however, that the detection (which is not necessarily equal to an accurate measurement) of an internal misaligned magnetic field in $\beta$~Cephei pulsators can be achieved by means of detecting additional peaks in multiplets.

\section*{Data availability}
The \ester models and mode asymmetries are available on Zenodo: \url{https://zenodo.org/records/17580179}

\begin{acknowledgements}
  We thank the anonymous referee for their comments on the manuscript, Dario Fritzewski for providing the distribution of fractions of critical rotation for the $\beta$~Cephei sample, and Zhao Guo for the discussions. 
  The research leading to these results has received funding from the European Research Council (ERC) under the Horizon Europe programme (Synergy Grant agreement N$^\circ$101071505: 4D-STAR). While partially funded by the European Union, views and opinions expressed are however those of the authors only and do not necessarily reflect those of the European Union or the European Research Council. Neither the European Union nor the granting authority can be held responsible for them. V.V. acknowledges support from the Research Foundation Flanders (FWO) under grant agreement N$^\circ$1156923N (PhD Fellowship). S.B.D. acknowledges funding from the European Union’s Horizon 2020 research and innovation programme under the Marie Skłodowska-Curie grant agreement N$^\circ$101034413. L.B. gratefully acknowledges support from the European Research Council (ERC) under the Horizon Europe programme (Calcifer; Starting Grant agreement N$^\circ$101165631). J.B., M.R., S.M. and J.S.G.M have been supported by CNES, focused on the preparation of the PLATO mission. Computations with \ester and \topc have made use of the HPC resources from the CALMIP supercomputing centre (Grant 2023-P0107). This research made use of the \texttt{numpy} \citep{Harris2020} and \texttt{matplotlib} \citep{Hunter2007} \texttt{Python} software packages.   
\end{acknowledgements}

%
%
\bibliographystyle{aa} 
\bibliography{main} 

\end{document}